\documentclass[aps,prl,preprint,showpacs,superscriptaddress,tightenlines]{revtex4}

\usepackage{graphicx}
\usepackage{dcolumn}
\usepackage{bm}
\usepackage{epsfig}

\begin{document}




\title{Measurement of Branching Fractions and 
Polarization in $B \to \phi K^{(*)}$ Decays}
       


\affiliation{Budker Institute of Nuclear Physics, Novosibirsk}
\affiliation{Chiba University, Chiba}
\affiliation{University of Cincinnati, Cincinnati, Ohio 45221}
\affiliation{University of Frankfurt, Frankfurt}
\affiliation{Gyeongsang National University, Chinju}
\affiliation{University of Hawaii, Honolulu, Hawaii 96822}
\affiliation{High Energy Accelerator Research Organization (KEK), Tsukuba}
\affiliation{Institute of High Energy Physics, Chinese Academy of Sciences, Beijing}
\affiliation{Institute of High Energy Physics, Vienna}
\affiliation{Institute for Theoretical and Experimental Physics, Moscow}
\affiliation{J. Stefan Institute, Ljubljana}
\affiliation{Kanagawa University, Yokohama}
\affiliation{Korea University, Seoul}
\affiliation{Kyungpook National University, Taegu}
\affiliation{Institut de Physique des Hautes \'Energies, Universit\'e de Lausanne, Lausanne}
\affiliation{University of Ljubljana, Ljubljana}
\affiliation{University of Maribor, Maribor}
\affiliation{University of Melbourne, Victoria}
\affiliation{Nagoya University, Nagoya}
\affiliation{Nara Women's University, Nara}
\affiliation{National Kaohsiung Normal University, Kaohsiung}
\affiliation{National Lien-Ho Institute of Technology, Miao Li}
\affiliation{Department of Physics, National Taiwan University, Taipei}
\affiliation{H. Niewodniczanski Institute of Nuclear Physics, Krakow}
\affiliation{Nihon Dental College, Niigata}
\affiliation{Niigata University, Niigata}
\affiliation{Osaka City University, Osaka}
\affiliation{Osaka University, Osaka}
\affiliation{Panjab University, Chandigarh}
\affiliation{Peking University, Beijing}
\affiliation{Princeton University, Princeton, New Jersey 08545}
\affiliation{RIKEN BNL Research Center, Upton, New York 11973}
\affiliation{University of Science and Technology of China, Hefei}
\affiliation{Seoul National University, Seoul}
\affiliation{Sungkyunkwan University, Suwon}
\affiliation{University of Sydney, Sydney NSW}
\affiliation{Toho University, Funabashi}
\affiliation{Tohoku Gakuin University, Tagajo}
\affiliation{Tohoku University, Sendai}
\affiliation{Department of Physics, University of Tokyo, Tokyo}
\affiliation{Tokyo Institute of Technology, Tokyo}
\affiliation{Tokyo Metropolitan University, Tokyo}
\affiliation{Tokyo University of Agriculture and Technology, Tokyo}
\affiliation{Toyama National College of Maritime Technology, Toyama}
\affiliation{University of Tsukuba, Tsukuba}
\affiliation{Utkal University, Bhubaneswer}
\affiliation{Virginia Polytechnic Institute and State University, Blacksburg, Virginia 24061}
\affiliation{Yokkaichi University, Yokkaichi}
\affiliation{Yonsei University, Seoul}

  \author{K.-F.~Chen}\affiliation{Department of Physics, National Taiwan University, Taipei} 
  \author{A.~Bozek}\affiliation{H. Niewodniczanski Institute of Nuclear Physics, Krakow} 
  \author{K.~Abe}\affiliation{High Energy Accelerator Research Organization (KEK), Tsukuba} 
  \author{K.~Abe}\affiliation{Tohoku Gakuin University, Tagajo} 
  \author{T.~Abe}\affiliation{High Energy Accelerator Research Organization (KEK), Tsukuba} 
  \author{I.~Adachi}\affiliation{High Energy Accelerator Research Organization (KEK), Tsukuba} 
  \author{H.~Aihara}\affiliation{Department of Physics, University of Tokyo, Tokyo} 
  \author{M.~Akatsu}\affiliation{Nagoya University, Nagoya} 
  \author{Y.~Asano}\affiliation{University of Tsukuba, Tsukuba} 
  \author{T.~Aso}\affiliation{Toyama National College of Maritime Technology, Toyama} 
% ...
\collaboration{The Belle Collaboration}

\begin{abstract}
We present the first measurement of decay amplitudes in 
$B \to \phi K^*$ and measurements of branching fractions in 
$B \to \phi K^{(*)}$ decays based on 
78.1~fb$^{-1}$ of data recorded at the $\Upsilon(4S)$ resonance with
the Belle detector at the KEKB $e^{+} e^{-}$ storage ring.
The decay amplitudes for the different $\phi K^{*0}$ helicity states  
are measured from the angular distributions of final state particles 
in the transversity basis. The longitudinal and transverse complex amplitudes
are $|A_0|^2 = 0.43 \pm 0.09 \pm 0.04$, 
$|A_\perp|^2 = 0.41 \pm 0.10 \pm 0.04$, 
$\arg(A_\parallel) = -2.57 \pm 0.39 \pm 0.09$, and
$\arg(A_\perp) = 0.48 \pm 0.32 \pm 0.06$.
The direct $CP$-violating asymmetries are found to be consistent with zero.
\end{abstract}

\maketitle



\def\mySpecialText{DRAFT v0.85 / \today}

\def\myspecial#1{}		     
\def\myspecial#1{\special{#1}}       



$B$ meson decays involving $b\to s\bar{s}s$ transitions, such as 
$B \to \phi K$ and $\phi K^*$, are forbidden to first order in the Standard Model (SM), 
but proceed by second order loop diagrams (penguin and box diagrams),
which lead to the flavor changing neutral current transition $b \to s$.
These processes provide information on the Cabibbo-Kobayashi-Maskawa matrix element $V_{ts}$ \cite{ref:ckm}
and are sensitive to physics beyond the SM
such as R-parity violating SUSY contributions to $b\to s\bar{s}s$ \cite{ref:datta}.
They can also be used to perform independent measurements of the 
$CP$-violating parameter sin$2\phi_{1}$ \cite{ref:fleischer}.
The branching fractions of $B \to \phi K$ have been 
predicted by QCD-factorization \cite{ref:QCDF} and PQCD \cite{ref:PQCD}.
The decay $B \to \phi K^*$ is a mixture of $CP$-even and $CP$-odd states;
polarization measurements allow us to project out the different $CP$ states statistically.

In this letter, we report the first measurement of the helicity state amplitudes in 
 $B^{0} \to \phi K^{*0}$ decay by a full three-dimensional angular analysis.
We also report measurements of branching fractions and direct $CP$ asymmetries
in  $B^+ \to \phi K^+$, $B^{0} \to \phi K^{0}$, 
$B^+ \to \phi K^{*+}$, and  $B^{0} \to \phi K^{*0}$ decays (charge
 conjugate modes are included everywhere unless otherwise specified).

This analysis is based on a data set with an integrated luminosity of
78.1~fb$^{-1}$ taken at the $\Upsilon(4S)$ resonance
recorded by the Belle detector \cite{ref:Belle}
at the KEKB  $e^{+}e^{-}$ collider \cite{ref:KEKB}.
This luminosity corresponds to $(85.5 \pm 0.5) \times 10^{6}$  
produced $B\bar{B}$ pairs. The beam energies are 8 GeV for $e^-$ and 3.5 GeV for $e^+$.

The Belle detector is a general purpose magnetic 
spectrometer equipped with a 1.5~T superconducting solenoid magnet. 
Charged tracks are reconstructed in a Central Drift Chamber (CDC)
and a Silicon Vertex Detector (SVD). 
Photons and electrons are identified using a CsI(Tl) Electromagnetic Calorimeter (ECL) 
located inside the magnet coil. 
Charged particles are identified using measured $dE/dx$
in the CDC as well as information from Aerogel Cherenkov 
Counters (ACC) and Time of Flight Counters (TOF). 
A kaon likelihood ratio, $P(K/\pi) = \mathcal{L}_K /(\mathcal{L}_K + \mathcal{L}_\pi)$, has values 
between 0 (likely to be a pion) and 1 (likely to be a kaon), where 
$\mathcal{L}_{K(\pi)}$ is derived from $dE/dx$, ACC and TOF measurements.


Candidate $\phi \to K^+K^-$ decays are found by selecting pairs of 
oppositely charged tracks that are not pion-like ($P(K/\pi)>0.1$).
The vertex of the candidate charged tracks is required to be consistent with 
the interaction point (IP) to suppress poorly measured tracks.
In addition, candidates are required to have a $K^+K^-$
invariant mass that is less than 10 MeV/$c^2$ from the nominal $\phi$ meson mass. 

Pairs of oppositely charged tracks are used
to reconstruct $K^0_S \to \pi^+\pi^-$ decays.
The $\pi^+\pi^-$ vertex is required to be displaced from 
the IP by a minimum transverse distance of 0.22~cm for
high momentum ($>1.5$ GeV/$c$) candidates and 0.08~cm for
those with momentum less than 1.5~GeV/$c$.
The direction of the pion pair momentum must agree with
the direction defined by the IP and the vertex displacement within 0.03 rad for 
high-momentum candidates, and within 0.1 rad for the remaining candidates.

Charged tracks with $P(K/\pi)>0.4$ ($<0.9$) are considered to be kaons (pions).
For $\pi^{0} \to \gamma\gamma$, a minimum photon energy of 50 MeV
is required and the $\gamma\gamma$ invariant mass must be less than 16 MeV/$c^2$
from the nominal $\pi^{0}$ mass. 
$K^{*}$ candidates are reconstructed in three decay modes: 
$K^{* 0} \to K^{+} \pi^{-}$, 
$K^{* +} \to K^{+} \pi^{0}$ 
and $K^{* +} \to K^{0}_{S} \pi^{+}$.
The invariant mass of the $K^{*}$ candidate is required to be less than 
70 MeV/$c^2$ from the nominal $K^{*}$ mass.

A $B$ meson is reconstructed from a $\phi$ meson candidate and 
a $K$ or $K^*$ candidate and identified by the
 energy difference 
$\Delta E \equiv E_B^{\rm cms} - E_{\rm beam}^{\rm cms}$, 
and the beam constrained mass $M_{\rm bc} \equiv \sqrt{(E^{\rm cms}_{\rm beam})^2 - (p_B^{\rm cms})^2}$.
$E_{\rm beam}^{\rm cms}$ is the beam energy in the center-of-mass system (cms) of the $\Upsilon(4S)$ resonance,
and $E_B^{\rm cms}$ and $p_B^{\rm cms}$ are the cms energy and momentum 
of the reconstructed $B$ candidate.
The $B$-meson signal window is defined as 5.27 GeV/$c^2$ $< M_{\rm bc} <$ 5.29 GeV/$c^2$
and $|\Delta E| <$ 64 (60) MeV for $B \to \phi K$ ($B \to \phi K^*$).
The signal window is enlarged to $-100$ MeV $<\Delta E<$ $80$ MeV for 
$B^+ \to \phi K^{*+} (K^{*+} \to K^{+} \pi^{0})$ 
because of the impact of shower leakage on $\Delta E$ resolution.
An additional requirement $\cos\theta_{K^*}<0.8$
is applied to reduce low momentum $\pi^0$ background,
where $\theta_{K^*}$ is the angle between the $K^*$ direction and 
its daughter kaon defined in the $K^*$ rest frame.
In the signal window about 1\% of the events have multiple candidates.
We choose the best candidate according to the value of the B vertex
$\chi^2$.


The dominant background is continuum 
$e^+e^- \to q\overline{q}$ production ($q=u,d,c,s$). 
Several variables 
are used to exploit the differences between 
the event shapes for continuum $q\overline{q}$ production
(jet-like) and for $B$ decay (spherical) in the cms frame of 
the $\Upsilon(4S)$ \cite{ref:continuum_suppression}.
These variables are combined into a single likelihood ratio 
$R_s = {\cal L}_s/({\cal L}_s + {\cal L}_{q\overline{q}})$, where 
${\cal L}_s$ (${\cal L}_{q\overline{q}}$) denotes the signal (continuum)
likelihood. 
An additional variable $\cos\theta_{H}$, which is the angle between
the $\phi$ momentum and the daughter kaon momentum 
in the $\phi$ rest frame, is included for the $\phi K^+$ and $\phi K^0_S$ channels. 

Backgrounds from other $B$ decay modes such as $B \to KKK^{(*)}$, 
$B \to f_0(980) K^{(*)}(f_0\to K^+K^-)$, $B \to \phi K\pi$, $B \to KKK\pi$, and 
feed-across between $\phi K^{*}$ and $\phi K$ decay channels are studied. 
The contributions from $B \to KKK^{(*)}$ and $B \to f_0(980) K^{(*)}(f_0\to K^+K^-)$ 
are estimated from the $K^+K^-$ invariant mass distribution.
The $K^+K^-$ mass distribution for $B \to KKK^{(*)}$ is determined by Monte Carlo (MC) simulation
assuming three-body phase space decay. 
The shape for $f_0(980)$ is obtained from MC, where a Breit-Wigner with a
40 MeV/$c^2$ intrinsic width is assumed.
These contributions are estimated separately by fits to the events outside of the $\phi$ mass region.
The contribution from $B \to KKK^{(*)}$ is estimated to be $5-9$\% \cite{ref:range} of the signal yield and
is subtracted from the raw signal yield.
The $B \to f_0 K^{(*)}$ contribution is estimated to be $2-12$\%.
The large uncertainty in the intrinsic width of the $f_0(980)$ is 
included in the systematic error.
The background from $B \to \phi K\pi$ decay, as well as 
higher $K^*$ resonance decay, is studied by performing fits to the $K\pi$ invariant mass.
The estimated background ($1-3$\%) is considered as a systematic error.
Contamination from four-body $B \to KKK\pi$ decays is checked by performing 
fits to the non-resonant region of $K^+K^-$ and $K\pi$ mass.
It is found to be very small and is neglected. 
The feed-across from $\phi K^{*}$ in $\phi K$ decay is removed 
by excluding events with $\Delta E < -120$ MeV from the fit. 
A veto is applied in $B\to \phi K^{*}$ channels to remove
the feed-across background from $\phi K$.

\begin{figure}[!htb]
\begin{center}
\resizebox*{3in}{4in}{\includegraphics{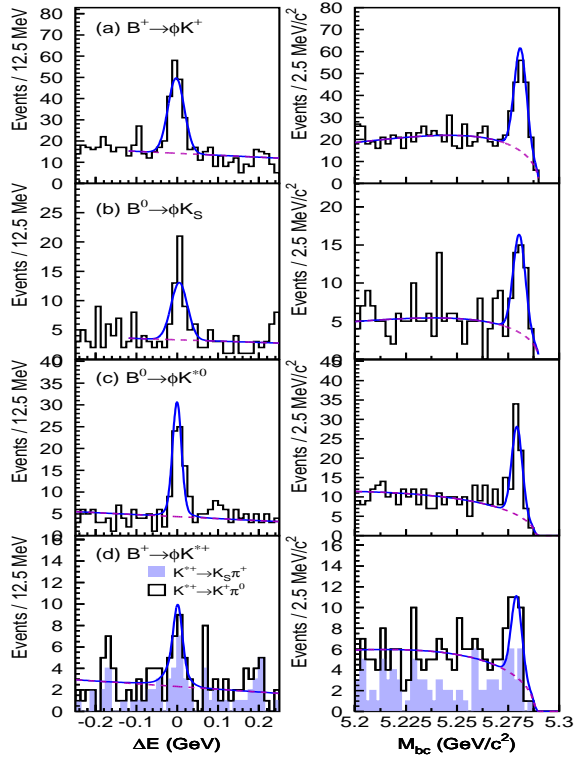}}
\end{center}
\caption{ Distributions of $\Delta E$ ($M_{\rm bc}$) with fit results for
the events in the $M_{bc}$ ($\Delta E$) signal window. 
The continuum background component is shown by dashed curves. }
\label{fig:demb-projections}
\end{figure}

The signal yields ($N_s$) are extracted by 
extended unbinned maximum-likelihood fits 
performed in $\Delta E$ and $M_{bc}$ simultaneously.
The signal probability density
functions (PDFs) are represented by Gaussians for both $\Delta E$ and $M_{\rm bc}$. 
The means and widths are 
verified using $B^{+}\to \overline{D^{0}}\pi^{+}$ and $B \to J/\psi K^*$ decays.
Additional bifurcated Gaussians (Gaussians with different widths 
on either side of the mean) are used to model 
the tails in the $\Delta E$ distribution of $\phi K^*$ channels.
The continuum PDF for $M_{\rm bc}$ ($\Delta E$) is determined from the events outside of 
$\Delta E$ ($M_{\rm bc}$) signal window.
The continuum PDFs for $M_{\rm bc}$ and $\Delta E$ are represented by 
an empirical background function introduced by ARGUS \cite{ref:argus}
and a linear function, respectively. 
The number of signal and background are floated in the fit while other PDF 
parameters are fixed.
The measured branching fractions (${\cal B}$) are summarized in Table~\ref{tab:bfs}.
The distributions of $\Delta E$ and $M_{\rm bc}$ for the four measured modes are shown in Fig.~\ref{fig:demb-projections}.


\begin{table}[!htb]
\caption{Signal yields ($N_s$) obtained by fits after background subtraction, 
total efficiency ($\epsilon$), statistical significance
 ($\Sigma \equiv \sqrt{2\ln[{\cal L}(N_s)/{{\cal L}(0)]}}$), and measured 
branching fraction ($\cal B$). The intermediate branching fractions are 
taken from \cite{ref:PDG}.} 
\label{tab:bfs}	 
\begin{tabular}{lccccc}
\hline 
\hline
Mode 			& $N_{s}$ 		  & $\epsilon$ (\%) & $\Sigma$ & ${\cal B}(10^{-6})$ \\
\hline
$B^{+} \to \phi K^{+}$	& $136^{+16}_{-15}$ & $16.9$ & $16.5$ & $9.4\pm 1.1\pm 0.7$ \\
\hline
$B^{0} \to \phi K^{0}$	& $35.6^{+8.4}_{-7.4}$ 	  & $4.6$  & $8.7$  & $9.0^{+2.2}_{-1.8}\pm0.7$ \\ 
\hline
$B^{0} \to \phi K^{*0}$ & $58.5^{+9.1}_{-8.1}$ & $6.9 $& $11.3$ & $10.0^{+1.6}_{-1.5}\;^{+0.7}_{-0.8}$ \\
\hline
$B^{+} \to \phi K^{*+}$	& $ - $                   & $ - $  & $4.9$  & $6.7^{+2.1}_{-1.9}\;^{+0.7}_{-1.0}$ \\
$~~~K^{*+}\to K^{+}\pi^{0}$	& $8.0^{+4.3}_{-3.5}$ 	  & $1.4$  & $2.8$  & $6.9^{+3.8}_{-3.2}\;^{+0.9}_{-1.0}$ \\
$~~~K^{*+}\to K^{0}_{S}\pi^{+}$& $11.3^{+4.5}_{-3.8}$   & $2.1$  & $4.0$  & $6.5^{+2.6}_{-2.3}\;^{+0.6}_{-0.9}$ \\
\hline 
\hline
\end{tabular}
\end{table}


The systematic errors in the signal yields are estimated 
by varying each fixed PDF parameter by $\pm 1\sigma$ of
its nominal value. Conservatively, the change in the signal yield from each variation 
is added in quadrature.  
The systematic errors in the efficiency are due to uncertainties in track finding (1\% per track),
particle identification (2\%), $K^{0}_{S}$ and $\pi^0$
finding (4\%), and to the uncertainty in ${\cal B}(\phi \to K^{+}K^{-})$ (1.4\%).
The estimated contaminations of $B\to f_0 K^{(*)}$ and $B\to \phi K\pi$ are included as an 
uncertainty in the background.
For the $B \to \phi K^{*}$ modes, an additional systematic error 
in the efficiency due to the uncertainty in the 
polarization together with the uncertainty in the slow pion detection efficiency (1\%-4\%) is included.


For the self-tagging modes $B^{+} \to \phi K^{+}$, $B^{0} \to \phi K^{*0}( K^+\pi^-)$, 
and $B^{+} \to \phi K^{*+}$ we have studied the direct $CP$ asymmetries
$A_{CP}=\frac{N(\overline{B} \to \overline{f})-N(B \to f)}
  {N(\overline{B} \to \overline{f})+N(B \to f)}$,
where $B$ ($\overline{B}$) is $B^0$ or $B^+$ ($\overline{B^0}$ or $B^-$) and
$f$ is one of the self-tagged $\phi K^{(*)}$ final states.
The values of $A_{CP}$ for $B^{+} \to \phi K^{+}$, $B^{0} \to \phi K^{*0}(K^+\pi^-)$, and 
$B^{+} \to \phi K^{*+}$ are: ${0.01 \pm 0.12 \pm 0.05}$, $0.07 \pm 0.15 ^{+0.05}_{-0.03}$,
and $-0.13 \pm 0.29^{+0.08}_{-0.11}$, respectively.
These correspond to 90\% confidence level limits of 
$-0.20 < A_{CP}(\phi K^{+}) < 0.22$, $-0.18 < A_{CP}(\phi K^{*0}(K^+\pi^-)) < 0.33$, 
and $-0.64 < A_{CP}(\phi K^{*+}) < 0.36$, respectively.
The systematic error includes the uncertainties in signal extraction (2\%) and 
detector induced bias ($1-6$\%),
which has been studied using large samples
of inclusive charged kaon and pion tracks, 
high momentum $D^{0} \to K^{-}\pi^{+}$ and $D^{+} \to K^{-}\pi^{+}\pi^{+}$ decays, and
$B$-meson decays to the channels $J/\psi K^{(*)}$, $\overline{D}^{0}\pi^{+}$ and
$D^{*-}\pi^{+}$. The systematic errors due to
background from $B \to f_0 K^{(*)}$ and non-$K^*$ background in $B\to \phi K^*$ channels 
are also included.

\begin{figure}[!htb]
\centerline{
\resizebox*{2.0in}{1.692in}{
\includegraphics{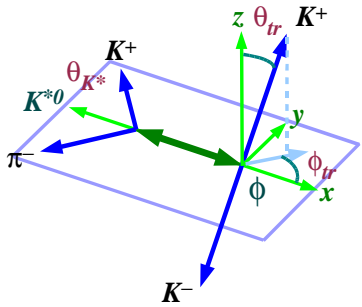}} }   
\caption{The definition of decay angles in $B \to \phi K^{*0}$ decay.}
\label{fig:angular-def}
\end{figure}

The decay angles of a $B$-meson, to the two vector mesons $\phi$ and $K^{*0}$, as defined
in the transversity basis \cite{ref:transversity}, are shown in Fig.~\ref{fig:angular-def}.
The $x$-$y$ plane is defined by the $K^{*0}$ daughters and the $x$ axis is in
the direction of the $\phi$-meson.
The $y$ axis is perpendicular to the $x$ axis and is on the same side as the kaon from $K^{*0}$ decay.
The $z$ axis is perpendicular to the $x$-$y$ plane according to the right-hand rule.
$\theta_{\rm tr}$ ($\phi_{\rm tr}$) is the polar (azimuthal) angle 
with respect to the z-axis of the $K^+$ from $\phi$ decay 
in the $\phi$ rest frame. $\theta_{K^*}$ is defined above.

The distribution of decays in the three angles \cite{ref:polarization_hepex}, $\theta_{K^*}$, $\theta_{\rm tr}$, and
$\phi_{\rm tr}$ is
\begin{eqnarray}
\label{equ:angularpdf}
\nonumber
&&{d^3 \Gamma (\phi_{\rm tr}, \cos\theta_{\rm tr}, \cos\theta_{K^*}) \over d\phi_{\rm tr} d\cos{\theta_{\rm tr}} d\cos{\theta_{K^*}}} 
= {9\over 32\pi} [ 
|A_\perp|^2 2 \cos^2{\theta_{\rm tr}} \sin^2{\theta_{K^*}} \\
\nonumber
&&~~~~~~ +|A_\parallel|^2 2 \sin^2{\theta_{\rm tr}} \sin^2{\phi_{\rm tr}} \sin^2{\theta_{K^*}} \\
\nonumber
&&~~~~~~ + |A_0|^2 4 \sin^2\theta_{\rm tr} \cos^2\phi_{\rm tr} \cos^2\theta_{K^*} \\
\nonumber
&&~~~~~~ +\sqrt{2} {\rm Re}(A^*_\parallel A_0) \sin^2\theta_{\rm tr}\sin 2 \phi_{\rm tr} \sin 2\theta_{K^*} \\
\nonumber
&&~~~~~~ -\eta\sqrt{2} {\rm Im}(A_0^* A_\perp) \sin 2\theta_{\rm tr} \cos\phi_{\rm tr} \sin 2\theta_{K^*} \\
&&~~~~~~ -2\eta{\rm Im}(A_\parallel^*A_\perp) \sin 2 \theta_{\rm tr} \sin \phi_{\rm tr} \sin^2 \theta_{K^*} ]~,
\end{eqnarray}
where $A_0$, $A_\parallel$, and $A_\perp$ are the complex amplitudes of the
three helicity states in the transversity basis with the normalization condition
$|A_0|^2 + |A_\parallel|^2 + |A_\perp|^2 = 1$, and $\eta\equiv +1$ ($-1$) for
$B^0$ ($\overline{B^0}$). $A_0$ denotes the longitudinal polarization of the $\phi\to K^+K^-$
system and $A_\perp$ ($A_\parallel$) is the transverse polarization
along the $z$-axis ($y$-axis).
The value of $|A_\perp|^2$ ($|A_0|^2 + |A_\parallel|^2 \equiv 1-|A_\perp|^2$)
is the $CP$-odd ($CP$-even)
fraction in the decay $B \to \phi K^*$ \cite{ref:polarization_hepex}. 
The imaginary phases of the amplitudes are sensitive to
final state interactions (FSI). 
The presence of FSI results in phases that are not either $0$ or $\pm\pi$.


The complex amplitudes are determined by performing
an unbinned maximum likelihood fit \cite{ref:j/psiK*polar} 
with $B^0 \to \phi K^{*0}(K^+\pi^-)$ candidates in the signal window.
The combined likelihood is given by
\begin{eqnarray}
\nonumber
&& \mathcal{L} =
\prod_i^N \epsilon(\theta_{K^*}, \theta_{\rm tr},\phi_{\rm tr}) [ f_{\phi K^{*0}}
\cdot\Gamma(\theta_{K^*}, \theta_{\rm tr},\phi_{\rm tr}) \\
\nonumber
&& ~~~~~~~~
+ f_{q\overline{q}}\cdot R_{q\overline{q}}(\theta_{K^*}, \theta_{\rm tr},\phi_{\rm tr})\\
&& ~~~~~~~~
+ f_{KKK^{*0}}\cdot R_{KKK^{*0}}(\theta_{K^*}, \theta_{\rm tr},\phi_{\rm tr}) ]~ ,
\end{eqnarray}
where $\Gamma$ is the angular distribution function (ADF) given by Eq.~\ref{equ:angularpdf}, and
$R_{q\overline{q}}$ and $R_{KKK^{*0}}$ are the
ADFs for continuum and $B \to KKK^{*0}$ background, respectively.
The value of $\eta$ is determined from the charge of the kaon in $K^{*0}$ decay,
$R_{q\overline{q}}$ is determined from sideband data and 
$R_{KKK^{*0}}$ is assumed to be flat. 
The detection efficiency function ($\epsilon$) is determined by MC.
The fractions of $\phi K^{*0}$ ($f_{\phi K^{*0}}$), $q\overline{q}$ ($f_{q\overline{q}}$) and
$B \to KKK^{*0}$ ($f_{KKK^{*0}}$) are parameterized as a function of $\Delta E$ and $M_{bc}$.
The value of $\arg(A_0)$ is set to zero and 
$|A_\parallel|^2$ is calculated from the normalization constraint in the fit.
Four parameters ($|A_0|^2$, $|A_\perp|^2$, $\arg(A_\parallel)$, and $\arg(A_\perp)$)
are left free to be determined from the fit. 

\begin{figure}[!htb]
\centerline{
\resizebox*{3.5in}{1.4in}{
\includegraphics{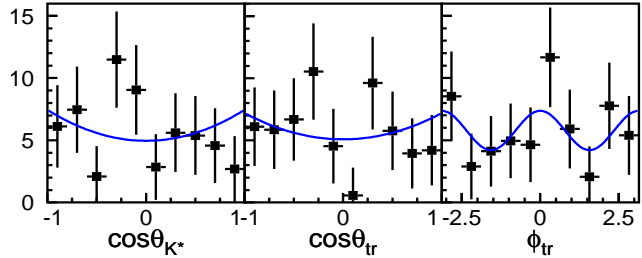}}
}   
\caption{ Projections of the transversity angles with results of the fit superimposed.
The points with error bars show the 
efficiency corrected data after subtraction of continuum and $B \to K^+K^-K^*$ background.
The $\chi^2$/n.d.f. for the projection of $\cos\theta_{K^*}$, 
$\cos\theta_{\rm tr}$ and $\phi_{\rm tr}$ are given by 1.05, 0.90 and 0.46, respectively.}
\label{fig:angular-proj}
\end{figure}

Figure~\ref{fig:angular-proj} shows projections
for each of the three angles together with results from the fit.
The amplitudes obtained from the fit are
$|A_0|^2 = 0.43 \pm 0.09 \pm 0.04$, 
$|A_\perp|^2 = 0.41 \pm 0.10 \pm 0.04$,
$\arg(A_\parallel) = -2.57 \pm 0.39 \pm 0.09$, and
$\arg(A_\perp) = 0.48 \pm 0.32 \pm 0.06$,
where the first errors are statistical and the second errors are systematic.
The systematic uncertainties 
include the  slow pion detection efficiency ($3-6$\%), the background
from higher $K^*$ states ($6-9$\%), and the $B\to f_0 K^*$ background ($1$\%). 
The systematic uncertainty due to the angular resolution is
estimated by MC simulation and found to be less than 1\%.
Uncertainties due to the background PDFs, the signal yields, and the modeling of 
efficiency function ($\epsilon$) are estimated to be $1-3$\%.


In summary, 
we measure the branching fractions of four 
$B \to \phi K^{(*)}$ decay modes.
The value of
${\cal B}(B^{+} \to \phi K^{+})$ is in good agreement
with, and supersedes, previously reported 
Belle measurements \cite{ref:conf07,ref:garmash}.
Our branching fraction results are in agreement with  
measurements by {\sc BaBar}~\cite{ref:babar2001} and CLEO~\cite{ref:cleo2001},
and the predictions by PQCD \cite{ref:PQCD}.
The measured direct $CP$ asymmetries in these modes are consistent with zero.
The decay amplitudes for $B^0 \to \phi K^{*0}$ are determined 
by fitting the angular distributions in the transversity basis.
The longitudinal polarization fraction ($f_L(\phi K^{*0})$) reported by 
{\sc BaBar} \cite{ref:babar_ex0303020} agrees with our measured value of $|A_0|^2$.  
The measured value of $|A_\perp|^2$ 
shows that both $CP$-odd ($|A_\perp|^2$) and $CP$-even ($|A_0|^2 + |A_\parallel|^2$) 
components are present in $\phi K^*$ decays,
in constrast to the case of $B \to J/\psi K^*$, which is dominantly $CP$-even \cite{ref:j/psiK*polar}. 
Our data also yield a good fit when the phases of
$A_\perp$ and $A_\parallel$ are constrained to zero and $-\pi$, 
indicating that our data cannot distinguish the presence of final state interactions.


We wish to thank the KEKB accelerator group for the excellent
operation of the KEKB accelerator.
We acknowledge support from the Ministry of Education,
Culture, Sports, Science, and Technology of Japan
and the Japan Society for the Promotion of Science;
the Australian Research Council
and the Australian Department of Industry, Science and Resources;
the National Science Foundation of China under contract No.~10175071;
the Department of Science and Technology of India;
the BK21 program of the Ministry of Education of Korea
and the CHEP SRC program of the Korea Science and Engineering Foundation;
the Polish State Committee for Scientific Research
under contract No.~2P03B 01324;
the Ministry of Science and Technology of the Russian Federation;
the Ministry of Education, Science and Sport of the Republic of Slovenia;
the National Science Council and the Ministry of Education of Taiwan;
and the U.S.\ Department of Energy.



\begin{thebibliography}{99}

\bibitem{ref:ckm}
{N. Cabibbo, Phys. Rev. Lett. {\bf 10}, 531 (1963);
 M. Kobayashi and T. Maskawa, Prog. Theor. Phys. {\bf 49}, 652 (1973).}

\bibitem{ref:datta}
{A. Datta, Phys. Rev. D {\bf 66} 071702, (2002).}

\bibitem{ref:fleischer}
{R. Fleischer, T. Mannel, Phys. Lett B {\bf 511}, 240 (2001).}

\bibitem{ref:QCDF}
{H.-Y. Cheng and K.-C Yang, Phys. Rev. D {\bf 64} (2001) 074004.}

\bibitem{ref:PQCD}
{C.-H. Chen, Y.-Y Keum, and H.-N. Li, Phys. Rev. D {\bf 64} (2001) 112002.}

\bibitem{ref:Belle}
{A. Abashian {\it et al.}, Nucl. Instr. Meth. A479, 117 (2002).}

\bibitem{ref:KEKB}
{S. Kurokawa, E. Kikutani, Nucl. Instr. Meth. A499, 1 (2003).}
   
\bibitem{ref:continuum_suppression}
{We use the $S_{\perp}$ variable as defined in
CLEO Collaboration, R. Ammar {\it et al.}, Phys. Rev. Lett. {\bf 71}, 674 (1993),
and the thrust angle and modified Fox-Wolfram
moments defined in Belle Collaboration, K. Abe {\it et al.}, Phys. Lett. B {\bf 517}, 309 (2001).}   

\bibitem{ref:range}
{The range corresponds to the range of values for different
sub-modes. This convention is used throughout this letter.}

\bibitem{ref:argus}
{The functional form is $x \sqrt{1-x^{2}}\exp(\alpha(1-x^{2}))$, where 
$x = {M_{bc}}/{E_{\rm beam}}$. 
ARGUS Collaboration, H. Albrecht {\it et al.} , Phys. Lett. B {\bf 241} (1990) 278; {\bf 254} (1991) 288.} 

\bibitem{ref:PDG}
{Particle Data Group, K. Hagiwara {\it et al.}, Phys. Rev. D {\bf 66} (2002).}

\bibitem{ref:transversity}
{I. Dunietz, H. Quinn, A. Snyder, W. Toki, and H.J. Lipkin, Phys. Rev. D {\bf 43}, 2193 (1991).}

\bibitem{ref:polarization_hepex}
{K. Abe, M. Satpathy and H. Yamamoto, hep-ex/0103002 (2001).}

\bibitem{ref:j/psiK*polar}
{Belle Collaboration, K. Abe {\it et al.}, Phys. Lett. B {\bf 538}, 11 (2002)}

\bibitem{ref:conf07}
{Belle Collaboration, H. Tajima, Int. J. Mod. Phys. A {\bf 17}, 2967-2981 (2002).}

\bibitem{ref:garmash}
{Belle Collaboration, A. Garmash {\it et al.}, Phys. Rev. D {\bf 65}, 092005 (2002).}   

\bibitem{ref:babar2001}
{{\sc BaBar} Collaboration, B. Aubert {\it et al.}, Phys. Rev. Lett. {\bf 87}, 151801 (2001) }

\bibitem{ref:cleo2001}
{CLEO Collaboration, R.A. Briere {\it et al.},  Phys. Rev. Lett. {\bf 86}, 3718 (2001).}     
     
\bibitem{ref:babar_ex0303020}
{{\sc BaBar} Collaboration, B. Aubert {\it et al.}, hep-ex/0303020.}     
     
\end{thebibliography}
\end{document}